\documentclass[epj,preprintnumbers,amsmath,amssymb]{svjour}
\usepackage{graphics, amssymb, amsmath}
\begin{document}
\title{Population dynamics on random networks: simulations and analytical models}
\author{Ganna Rozhnova \and Ana Nunes
}                     
%
%
\institute{Centro de F{\'\i}sica da Mat{\'e}ria Condensada and
Departamento de F{\'\i}sica,\\
Faculdade de Ci{\^e}ncias da Universidade de
Lisboa,\\
P-1649-003 Lisboa Codex, Portugal
}
\date{Received: date / Revised version: date}
%
\abstract{
We study the phase diagram of the standard pair approximation equations for two different models in population dynamics, the susceptible-infective-recovered-susceptible model of infection spread and a predator-prey interaction model, on a network of homogeneous degree $k$. These models have similar phase diagrams and represent two classes of systems for which noisy oscillations, still largely unexplained, are observed in nature. We show that for a certain range of the parameter $k$ both models exhibit an oscillatory phase in a region of parameter space that corresponds to weak driving. This oscillatory phase, however, disappears when $k$ is large. For $k=3, 4$, we compare the phase diagram of the standard pair approximation equations of both models with the results of simulations on regular random graphs of the same degree. We show that for parameter values in the oscillatory phase, and even for large system sizes, the simulations either die out or exhibit damped oscillations, depending on the initial conditions. We discuss this failure of the standard pair approximation model to capture even the qualitative behavior of the simulations on large regular random graphs and the relevance of the oscillatory phase in the pair approximation diagrams to explain the cycling behavior found in real populations.
} 
\maketitle
\section{Introduction}
\label{intro}
A common paradigm in population dynamics studies is to assume that population is not spatially distributed so that individuals mix perfectly and contact each other with equal probability. Thus, in the limit of infinite population the time evolution of the system is described in terms of the densities of different subpopulations and governed by a set of differential equations which can be deduced from the law of mass action \cite{murray,anderson_may}. Another approach is to use stochastic dynamics on lattices or on more general graphs where the discrete variables associated to each node represent the presence and attributes of an individual, e.g. susceptible, infective or recovered state in the case of epidemic dynamics \cite{contact_rev1,contact_rev2}. These studies have shown that spatial correlations neglected in the perfect mixing assumption play an important role in the behavior of population dynamics on graphs, and therefore also in real populations. 

In order to try to capture the effects of spatial correlations in the framework of an analytic description the standard pair approximation (PA) as well as several improvements based on including higher order correlations or on different closure assumptions at the level of pairs have been proposed in the context of ecological and epidemiological models \cite{ref_PA1,ref_PA2,ref_PA3,ref_PA4,pre,rand_morris_jerome1,rand_morris_jerome2,rand_morris_jerome3,proc,lebowitz,tome1,tome2,nikoloski}. The performance of the PA was  compared with the perfect mixing or mean field approximation (MFA) for several different population dynamics models on lattices and other graphs \cite{lebowitz,tome1,tome2,nikoloski,volz}. In these studies, the PA was found to perform much better than the MFA in the prediction of the steady state of the system, and to give a good quantitative agreement in a large region of parameter space.
On the other hand, a recent study of an epidemiological model has shown that 
an overall accurate prediction may require going beyond the PA to higher order cluster approximations \cite{ourpre2}.

While being more accurate than the MFA, the PA equations are still simple enough to allow for a complete description of their 
phase diagram, which is typically richer than the MFA phase diagram. In particular, models based on 
susceptible-infective-recovered-susceptible (SIRS) infection dynamics on lattices have been shown 
to exhibit in the PA a region in parameter space corresponding to stable cycles \cite{rand_morris_jerome1,rand_morris_jerome2,rand_morris_jerome3}. This region is small enough to be easily missed in a coarse grained numerical study but it is large in the admissible parameter region of an important class of diseases for which the susceptible turn over rate is much slower than the recovery rate (the weak driving regime). In particular, this oscillatory phase corresponds to realistic parameter estimations for childhood infectious diseases and could be related with the phenomenon of non-seasonal recurrent epidemics found in different data records in the pre-vaccination period for this class of infections \cite{anderson_may,pre,bauchearn}. 

Predator-prey systems are another class of population dynamics where non-seasonal cycling behavior has been reported for long \cite{elton,exp_populations,lynxhare}, and robust simple models exhibiting unforced oscillations have been much sought for \cite{murray,tome1,tome2,lotka,volterra,lipowski,droz,mobilia,alava}. In this case also, 
sustained oscillations have been identified in a small region of the phase diagram of the PA equations of a two parameter predator-prey model on the square lattice \cite{tome1,tome2}. Not much is known about predator-prey interaction parameters \cite{predpreyparameters}, and this small oscillatory phase may correspond to biologically realistic parameter values for some predator-prey pairs.

In this paper we consider these two paradigms in population dynamics: an epidemic model (Section \ref{sirs}) and a predator-prey model (Section \ref{abes}). We discuss the relevance of the oscillatory phase in the PA phase diagrams to explain the cycling behavior found in nature by comparing the behavior of the PA deterministic equations in the oscillatory phase with the results of simulations on regular random graphs (RRGs) of different degrees for both systems. The PA is exact on the Bette lattice, and RRGs are locally tree-like, so one would expect stochastic dynamics running on RRGs to be well approximated by the PA equations, at least for large system sizes. However, we found that, in both cases, the oscillations predicted by the PA are suppressed in the simulations, even for system sizes as large as $N=5\times 10^7$ nodes. We expect that simulations on more realistic networks, where the PA performs worse than for RRGs, will also fail to exhibit oscillations. In this sense, the oscillatory phases are an artifact of the PA because they disappear in finite networks. In Section \ref{discussion} we discuss the reasons why the PA fails even qualitatively to describe the behavior of these systems and sum up the conclusions of the present study.

\section{The epidemic model}
\label{sirs}
In this section, we consider the dynamics of the SIRS epidemic model on a random network of homogeneous degree $k$ and $N$ nodes, a regular random graph of degree $k$ (RRG-$k$). A preliminary report on the results of this section can be found in Ref. \cite{proc}. In the SIRS model each node can be occupied by an individual in susceptible ($S$), infected ($I$), or recovered ($R$) state and the following update rules with asynchronous update are applied. Infected individuals recover at rate $\delta$, recovered individuals lose immunity at rate $\gamma$, and infection of a susceptible node occurs at infection rate $\lambda$ multiplied by the number of its infected nearest neighbors $n \in \{0,1,\ldots,k\}$:
\begin{eqnarray}
\label{sirs_rates} 
I\stackrel{\delta}{\rightarrow}R \ ,  \\
R\stackrel{\gamma}{\rightarrow}S  \ , \nonumber\\
S\stackrel{\lambda \cdot n}{\rightarrow} I \ . \nonumber
\end{eqnarray}

In the infinite population limit, with the assumptions of spatial homogeneity and uncorrelated pairs, the system is described by the deterministic
\begin{figure}
\centering
\resizebox{1.0\columnwidth}{!}{%
  \includegraphics{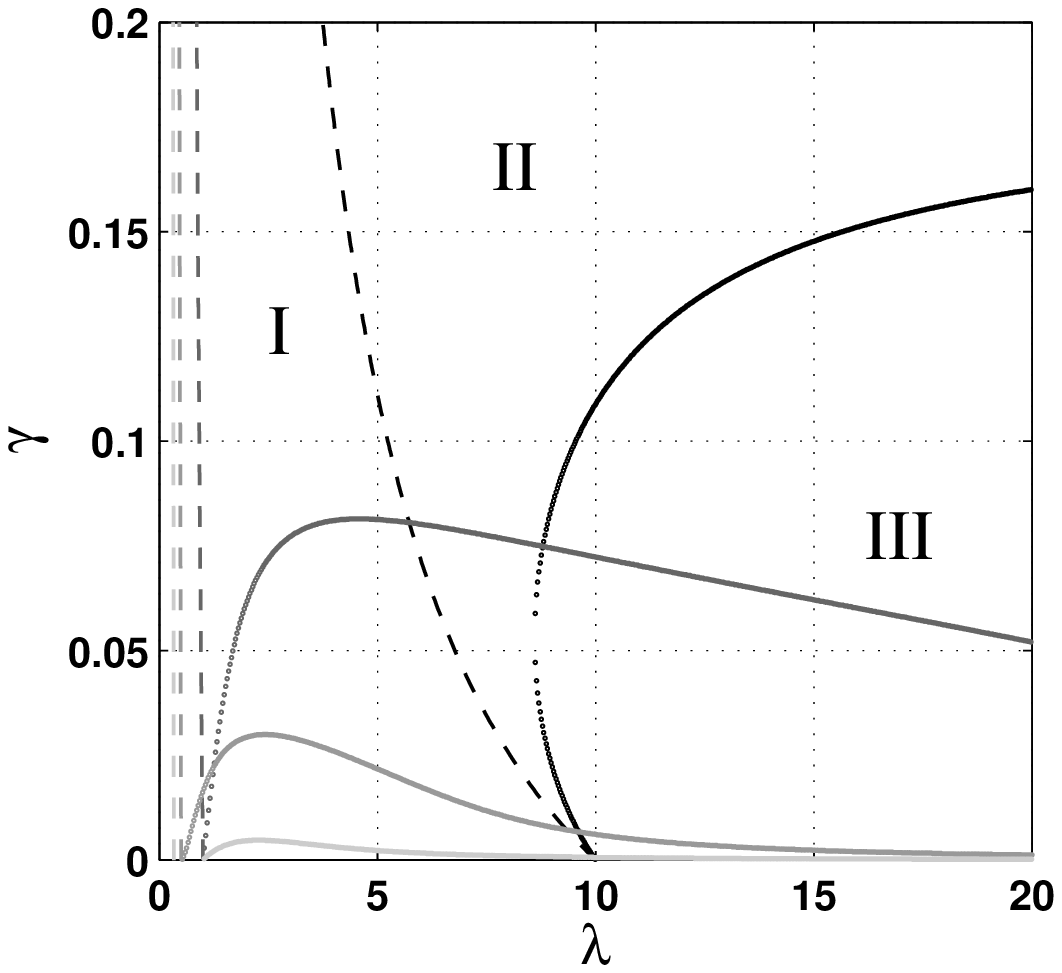}
  }
\caption{Phase
diagram in the $(\lambda,\gamma)$ plane of the PA SIRS model for $\delta=1$ and $k=2.1, 3, 4, 5$ (darker shades of gray for smaller values of $k$). Region I (regions II and III) represents susceptible-absorbing states (active states with nonzero infective and susceptible densities). The dashed lines correspond to the transcritical bifurcation curves, see Eq. (\ref{lambdac}); the dotted lines are associated with a supercritical Andronov-Hopf bifurcation. In region II the fixed points are asymptotically stable nodes or asymptotically stable foci, in region III asymptotically stable solutions are limit cycles. The markers depicting different phases (I, II, III) are shown for $k=2.1$.}
\label{bif_diagrams}
\end{figure}
equations of the standard or uncorrelated PA \cite{pre,lebowitz}: 
\begin{eqnarray}
\label{paeq} 
\frac{dP_S}{dt}&=&\gamma(1-P_S-P_I)-k\lambda P_{SI} \ ,\\
\frac{dP_I}{dt}&=&k\lambda P_{SI}-\delta P_I \ , \nonumber\\ 
\frac{dP_{SI}}{dt}&=&\gamma P_{RI} + \frac{(k-1)\lambda P_{SI}}{P_S}(P_S-P_{SR}-2P_{SI})-\nonumber\\
                               &-&(\lambda+\delta) P_{SI} \ , \nonumber\\
\frac{dP_{SR}}{dt}&=&\delta P_{SI}+\gamma (1-P_{S}-P_I-P_{RI}-2P_{SR})-\nonumber\\
                  &-&\frac{(k-1)\lambda P_{SI}P_{SR}}{P_S} \ , \nonumber\\
\frac{dP_{RI}}{dt}&=&\delta (P_I-P_{SI})-(\gamma+2\delta) P_{RI}+\frac{(k-1)\lambda P_{SI} P_{SR}}{P_S} \ . \nonumber
\end{eqnarray}
In the above equations, the variables $P_S$, $P_I$ stand for the probability that a randomly chosen node is in state $S$, $I$, and the variables $P_{SI}$, $P_{SR}$, $P_{RI}$ stand for the probability that a randomly chosen pair of nearest neighbor nodes is an $SI$, $SR$, $RI$ pair, as $N\to\infty$. 
As expected, neglecting the pair correlations and setting the pair state probabilities equal to the product of the node state probabilities these equations reduce to the classic equations of the perfectly mixed or MFA SIRS model \cite{anderson_may}. 

The phase diagram of the PA SIRS deterministic model [Eq. (\ref{paeq})] for $k=4$ was analyzed in Ref. \cite{pre}. The phase diagram of the same model 
for several values of $k$ in the range $k>2$ is plotted in Fig. \ref{bif_diagrams}, where we have set the time scale so that $\delta =1$. The shades of gray of the corresponding critical curves are darker for decreasing $k$. In the following discussion we will refer to the markers depicting different phases (I, II, III) which are shown for $k=2.1$ in Fig. \ref{bif_diagrams}. Region I (regions II and III)  represents susceptible-absorbing (active) states. The critical lines separating the absorbing and the active phases (the dashed lines) correspond to the transcritical bifurcation curves and are 
\begin{figure*}
\centering
\resizebox{0.75\textwidth}{!}{%
  \includegraphics{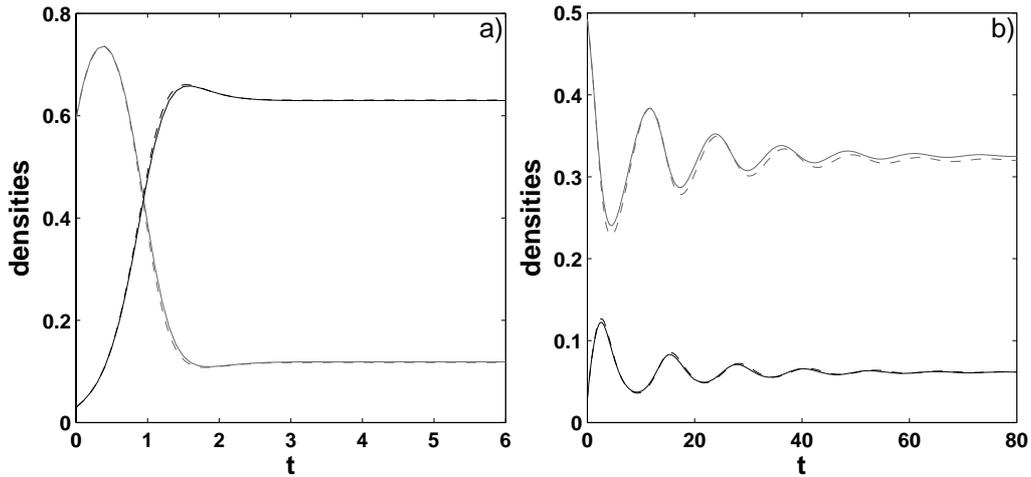}
  }
\caption{For $\delta=1$, $k=4$, comparison of the solutions of the PA deterministic model (dashed lines) with the results of stochastic simulations (solid lines) on a RRG-$4$ with $N=10^6$ nodes for parameter values in region II. Susceptible (infective) densities are plotted in gray (black). Parameters: (a) $\gamma=2.5$, $\lambda=2.5$; (b) $\gamma=0.1$, $\lambda=2.5$.}
\label{timeseries}
\end{figure*}
given by: 
\begin{equation}
\label{lambdac}
\lambda_{c}(\gamma)=\frac{\gamma +1}{(k-1)\gamma+k-2} \ .
\end{equation}
Within the active phase, the fixed points can be asymptotically stable nodes or asymptotically stable foci as in region II, in region III asymptotically stable solutions are limit cycles. The critical curves corresponding to a supercritical Andronov-Hopf bifurcation of the nontrivial equilibrium (the dotted lines) separate the active phase with constant densities from an active phase with stable oscillatory behavior. The oscillatory phase is large for 
$k\gtrsim 2$ and it gets thinner as $k$ increases, but it persists in the whole range $2< k \lesssim 6$. 

A similar phase diagram, with the Andronov-Hopf bifurcation critical line bounding an oscillatory phase, was reported in other studies of related models on lattices \cite{rand_morris_jerome1,rand_morris_jerome2,rand_morris_jerome3}, where SIR dynamics with different mechanisms of replenishment of susceptibles is modelled at the level of pairs with the standard PA or another closure approximation. All these different models exhibit an oscillatory phase in the regime of weak driving through slow introduction of new susceptible individuals (small $\gamma $ in the present case). 

The analysis of the SIRS model with the update rules given by Eq. (\ref{sirs_rates}) was performed by using both the PA and stochastic simulations on the square lattice in Ref. \cite{lebowitz}. However, the critical line bounding an oscillatory phase seems to have been missed in the previous study, and for parameter values in the oscillatory phase no global oscillations were found in the simulations on the square lattice. 

We have compared the behavior of the PA SIRS model [Eq. (\ref{paeq})] for $k=3, 4$ with the results of stochastic simulations on RRGs for several system sizes. In the stochastic simulations, the system was set in a random initial condition with given node and pair densities, and an efficient algorithm for stochastic processes in spatially structured systems \cite{gillespie1} was implemented to update the states of the nodes according to the processes of infection, recovery and immunity waning, see Eq. (\ref{sirs_rates}). In the nonspatial case, this algorithm reduces to the well-known Gillespie's method \cite{gillespie2}. Unless stated otherwise, the results were obtained for RRGs with $N=10^6$ nodes. In the surviving stochastic simulations, the system always approaches a well-defined nontrivial equilibrium in an over-damped or under-damped way, exhibiting small fluctuations around the average trajectory. To compare with the solutions of the deterministic model, for each set of parameter values and initial conditions, the simulations were averaged over $10^3$ realizations of a RRG.  

\begin{figure*}[t]
\centering
\resizebox{0.75\textwidth}{!}{%
  \includegraphics{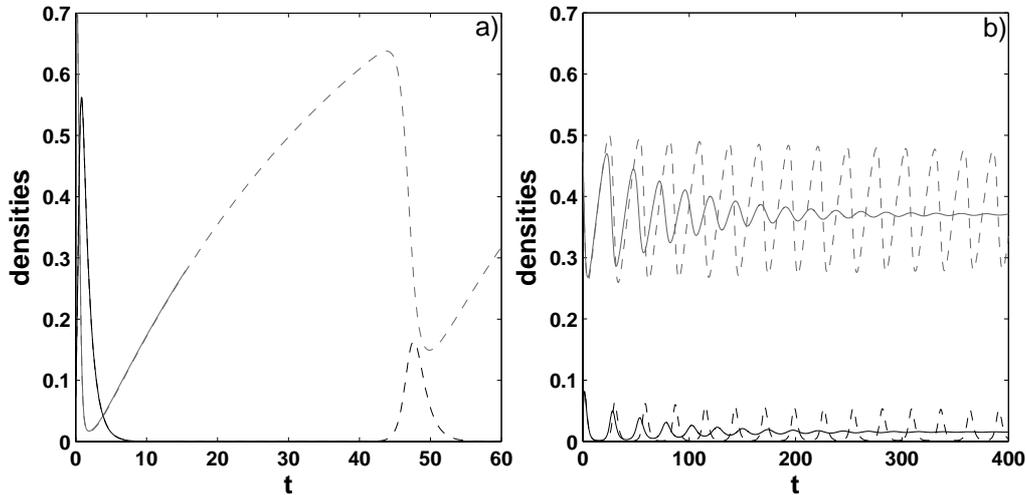}
  }
\caption{For $\delta=1$, $k=4$, $\gamma=0.025$, $\lambda=2.5$, comparison of the solutions of the PA deterministic model (dashed lines) with the results of stochastic simulations (solid lines) on a RRG-$4$ with $N=10^6$ nodes for parameter values in region III. Susceptible (infective) densities are plotted in gray (black). Initial conditions: (a) $P_S(0)\approx0.9240$, $P_I(0)\approx0.0731$, $P_{SI}(0)\approx0.0558$, $P_{SR}(0)\approx0.0024$, $P_{RI}(0)\approx0.0005$; (b) $P_S(0)\approx0.4889$, $P_I(0)\approx0.0287$, $P_{SI}(0)\approx0.0104$, $P_{SR}(0)\approx0.2369$, $P_{RI}(0)\approx0.0129$.}
\label{cycles}
\end{figure*}

The results of some of these stochastic simulations on a RRG-$4$ and solutions of the PA SIRS equations [Eq. (\ref{paeq})] are shown in Fig. \ref{timeseries} (region II) and in Fig. \ref{cycles} (region III). The susceptible (gray lines) and the infective (black lines) densities are shown in  Fig. \ref{timeseries} for two sets of parameter values with fixed infection rate $\lambda$ and decreasing rate of immunity waning $\gamma$. The numerical solutions of the PA SIRS equations are plotted in dashed lines, and the results of the simulations in solid lines. For parameter values well within region II of the phase diagram as in Fig. \ref{timeseries}(a) there is excellent agreement between the solutions of the PA SIRS model for the same initial densities and the results of the stochastic simulations, both for the transient behavior and for the steady states. This agreement deteriorates as $\gamma$ decreases and the boundary of the oscillatory region is approached, see Fig. \ref{timeseries}(b).  

The susceptible (gray lines) and the infective (black lines) densities are shown in Fig. \ref{cycles} for two sets of initial conditions for parameter values in the oscillatory region III. Most simulations (solid lines) die out after a short transient, see Fig. \ref{cycles}(a), while the corresponding solutions of the PA SIRS deterministic model (dashed lines) converge to the stable limit cycle for all initial conditions (a typical set is chosen in the plot). By choosing initial conditions not far from the stable cycle predicted by the PA SIRS model to avoid extreme susceptible depletion during the transient, damped oscillations towards a nontrivial equilibrium may also be observed in region III. In Fig. \ref{cycles}(b) a plot is shown of one of these surviving simulations (solid lines), together with the solution of the PA equations (dashed lines) for the same parameter values and initial
conditions. Thus, instead of an oscillatory phase, the stochastic model on a RRG-$4$ exhibits in region III a bistability phase, even for large system sizes.

As is seen in Fig. \ref{bif_diagrams}, the oscillatory phase for the PA SIRS deterministic model for $k=3$ is large in comparison with that for $k=4$. It is easier in this case to  avoid stochastic extinctions and to compare with the numerical simulations on a RRG-$3$ for a broad range of parameter values. Nevertheless, we have found that in the endemic phase of the phase diagram the stochastic dynamics of the SIRS on a RRG-$3$ is qualitatively 
the same as on a RRG-$4$. For large $\gamma$ it is well approximated by the PA deterministic equations but no global oscillations are observed for small $\gamma$. 

This failure of the PA model to capture the qualitative behavior of the simulations on large RRGs has been investigated. Extinctions due to finite size are one of the reasons why the oscillatory phase is seen as an absorbing phase in the stochastic simulations. Indeed, as can be seen in Fig. \ref{cycles}(a), the oscillations predicted by the PA SIRS deterministic model attain
very small densities of infectives during a significant fraction of the period in the transient regime. For example, the density of infectives attains values smaller than $10^{-5}$ in the transient regime for the initial conditions of Fig. \ref{cycles}(a). This problem can be overcome either by setting a RRG in a random initial condition with node densities lying very close to the limit cycle and/or by increasing system size. Though at low rate of immunity waning the SIRS dynamics on RRGs is blurred by strong fluctuations and stochastic extinctions, it is quite surprising that global oscillatory behavior has not been identified in stochastic simulations at all. Increasing system size up to $N=5\times 10^7$ nodes we still find suppression of oscillations in region III and significant discrepancies between the transient and steady states of the PA SIRS solutions and the results of the simulations in region II close to the boundary with region III.

\section{The predator-prey model}
\label{abes}

A small oscillatory phase has also been identified in the phase diagram of the PA equations of a two parameter predator-prey model on the square lattice \cite{tome1,tome2}.
In this section we consider a generalization of this model and study whether the oscillations predicted by the PA equations persist in stochastic simulations on RRGs. 

Consider a predator-prey model in which each node of a RRG-$k$ can be either empty ($E$) or occupied by a predator ($A$) or a prey ($B$). Let $n_1, n_2 \in \{0,1,\ldots,k\}$ denote the number of nearest neighbors occupied by a prey in the neighborhood of a node in state $E$ and by a predator in the neighborhood of a node in state $B$, respectively. The dynamics of the system at the microscopic level is governed by the following four processes: death of predators with rate $d$, birth of prey with rate $b\cdot n_1$, and two competing predator-prey interactions with rates $p_1\cdot n_2$ and $p_2\cdot n_2$ associated with predator reproduction and consumption of prey. 
\begin{figure*}
\centering
\resizebox{0.75\textwidth}{!}{%
  \includegraphics{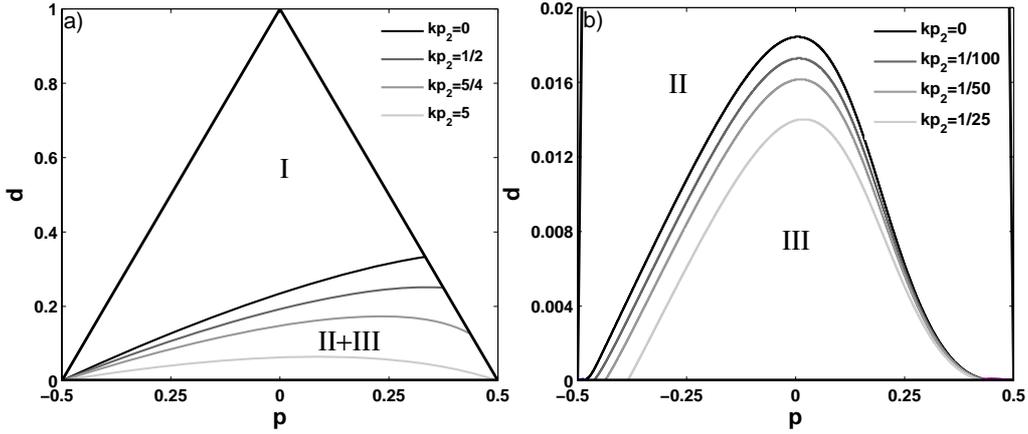}
  }
\caption{Phase diagram for the predator-prey model described by Eq. (\ref{abeeq}) for $k=4$. The constraints on the independent parameters $p$ and $d$ given by Eq. (\ref{konstr}) define the triangle area as parameter space for fixed $k\cdot p_2$. (a) Region I represents prey-absorbing states and regions II and III represent active states with nonzero densities of predators and prey. Several critical lines of transcritical bifurcations between prey-absorbing and active phases are shown for $k=4$ and different values of $p_2$, see Eq. (\ref{kvad}). (b) The active states can be asymptotically stable nodes or asymptotically stable foci as in region II or stable limit cycles as in region III. The phase with stable oscillatory behavior is bounded by supercritical Andronov-Hopf bifurcation curve plotted for $k=4$ and different values of $p_2$ for comparison.}    
\label{abe}
\end{figure*}
The state of the system evolves in time with asynchronous update according to the following set of local rules:
\begin{eqnarray}
\label{pa_rates} 
A\stackrel{d}{\rightarrow}E \ , \\
E\stackrel{b\cdot n_1}{\rightarrow}B  \ ,\nonumber \\
B\stackrel{p_1 \cdot n_2}{\rightarrow} A \ , \nonumber\\
B\stackrel{p_2 \cdot n_2}{\rightarrow} E \ . \nonumber
\end{eqnarray}

The standard PA model for this dynamics on a RRG of coordination number $k$ is described by:
\begin{eqnarray}
\label{abeeq}
\frac{dP_A}{dt}&=&-dP_A+kp_1P_{AB} \ , \\
\frac{dP_B}{dt}&=&-k(p_1+p_2)P_{AB}+kbP_{BE} \ , \nonumber\\ 
\frac{dP_{AB}}{dt}&=&-(p_1+p_2+d)P_{AB}+\frac{(k-1)bP_{AE}P_{BE}}{1-P_A-P_B}+\nonumber\\
                  &+&\frac{(k-1)P_{AB}}{P_B}\left[p_1(P_B-2P_{AB}-P_{BE})-p_2P_{AB}\right] \ , \nonumber\\
\frac{dP_{AE}}{dt}&=&d(P_A-P_{AB}-2P_{AE})-\frac{(k-1)bP_{AE}P_{BE}}{1-P_A-P_B}+\nonumber\\
                  &+&p_2P_{AB}+\frac{(k-1)P_{AB}}{P_B}(p_1P_{BE}+p_2P_{AB}) \ , \nonumber\\
\frac{dP_{BE}}{dt}&=&\frac{(k-1)P_{AB}}{P_B}\left[p_2(P_B-P_{AB}-2P_{BE})-p_1P_{BE}\right]+\nonumber\\
 &+&\frac{(k-1)bP_{BE}}{1-P_A-P_B}(1-P_A-P_B-P_{AE}-2P_{BE})-  \nonumber\\
                  &-&bP_{BE}+dP_{AB} \ , \nonumber
\end{eqnarray}
where, using the notation of the previous section, the variables stand for the limit values of the node $P_A$, $P_B$ and pair densities $P_{AB}$,
$P_{AE}$, $P_{BE}$ as $N\to\infty$. 

Linear stability analysis of the equilibrium points of Eq. (\ref{abeeq}) allows to identify different regions in the phase diagram, whose parameters are the coordination number of a RRG and the four rate constants of the model. For $p_2=0$, Eq. (\ref{abeeq}) reduces to the model studied in Ref. \cite{tome1}. Following \cite{tome1}, the number of independent parameters can be reduced by taking  
\begin{equation}
kp_1=\left(\frac{1}{2}+p-\frac{d}{2}\right) \ , \ kb=\left(\frac{1}{2}-p-\frac{d}{2}\right) \ ,
\end{equation}
so that $k(p_1+b)+d=1$, leaving two independent parameters $p$ and $d$ with the restrictions 
\begin{equation}
\label{konstr}
-1/2\leq p \leq 1/2 \ , \ 0\leq d\leq 1 \ . 
\end{equation}
For fixed $k$, the parameter space is then given by a right triangular prism in which the base is restricted to the allowed values of $(p,d)$ and the height is defined by the parameter $k \cdot p_2\geq 0$. Equivalently, the phase diagram can be plotted in the $(p,d)$ plane for the intersections $k\cdot p_2=const$ of the prism. 

\begin{figure*}
\centering
\resizebox{0.75\textwidth}{!}{%
  \includegraphics{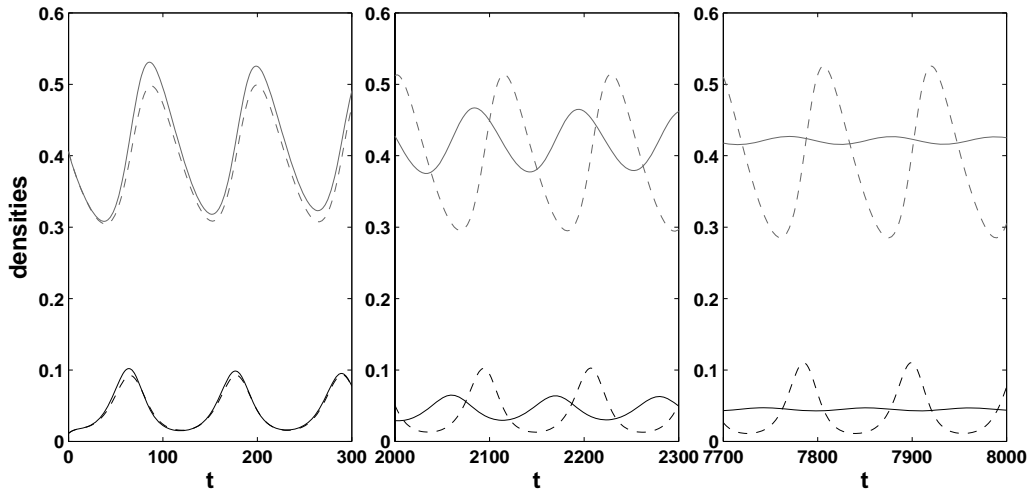}
  }
\caption{For $k=4$, $p=0$, $d=0.017$, $p_2=0$, comparison of the solutions of the PA deterministic model given by Eq. (\ref{abeeq}) (dashed lines) with the results of stochastic simulations (solid lines) on a RRG-$4$ with $N=10^6$ nodes for parameter values in region III. Predator (prey) densities are plotted in gray (black). The three panels show time interval of equal length in the beginning, in the middle and in the end of the same time series.}
\label{timeseries_abe}
\end{figure*} 

In Fig. \ref{abe} we have analyzed the phase diagram of the predator-prey model described by Eq. (\ref{abeeq}) for $k=4$. In Fig. \ref{abe}(a) region I represents prey-absorbing states in which a trivial fixed point is stable, regions II and III correspond to active states containing nonzero predator and prey densities. For $k=4$ and fixed $p_2$, the critical lines $d_c=f(p)$ separating prey-absorbing states from active states correspond to the transcritical bifurcation curves that are given by the positive root of the quadratic equation: 
\begin{equation}
\label{kvad}
kp_2=\frac{3-6d_c-29d_c^2+32d_cp-12p^2}{8d_c} \ . 
\end{equation}
Several solutions of Eq. (\ref{kvad}) for $k=4$ and different values of $p_2$ are shown in Fig. \ref{abe}(a). As for the asymptotic behavior of the system within the active phase, Fig. \ref{abe}(b) shows two different regions: region II correponds to asymptotically stable nodes or asymptotically stable foci (not separated in the plot) and region III represents stable limit cycles. The stability of the fixed point in region II is lost on the critical line separating regions II and III through a supercritical Andronov-Hopf bifurcation where a limit cycle is born. Again, several critical curves for $k=4$ and different values of $p_2$ are plotted in Fig. \ref{abe}(b). Note that the oscillatory phase occuring for small values of $d$ is also present for other sets of parameters, in particular for $k\neq4$, but in general it is small when compared with the whole parameter space (see the scale in the left and in the right panels in Fig. \ref{abe} for comparison). For fixed $p_2$, the dependence of the oscillatory phase on $k$ is similar to that of the SIRS model. Moreover, the oscillatory phase is robust with respect to several perturbations of the model (for example, with respect to the inclusion of the process describing natural death of prey). The analysis of a simplified model with $p_2=0$ and $k=4$ was performed by using both the PA and stochastic simulations on the square lattice \cite{tome1}. For parameter values in the oscillatory phase, no global oscillations were found in the simulations. More surprisingly, we have found that the oscillations are also suppressed in simulations on a RRG-$4$, even for system sizes as large as $N=5\times 10^7$ nodes.

The typical results of the averaged stochastic simulations on a RRG-$4$ and solutions of the PA model given by Eq. (\ref{abeeq}) for the same initial conditions and parameter values in the oscillatory region III are shown in Fig. \ref{timeseries_abe}. The figure is composed by three panels showing time intervals of equal length in the beginning, in the middle and in the end of a long time series. The predator (prey) densities are plotted in gray (black). The numerical solutions of the PA equations (dashed lines) converge to a limit cycle but the results of the simulations (solid lines) exhibit slowly damped oscillations towards a nontrivial equilibrium.
   

\section{Discussion and conclusions}
\label{discussion}

The breakdown of the PA in the oscillatory phase found for both the epidemic and the predator-prey model is partially due to stochastic extinctions in the simulations. However, by choosing carefully the initial conditions, the simulations persist for long times exhibiting damped oscillatory behavior towards a nontrivial equilibrium, which shows that there are other effects at play. 

We shall discuss these in the framework of the SIRS model, which is simpler, although a similar reasoning applies to the predator-prey model.  As pointed out in Section \ref{sirs}, a crucial parameter for the performance of the PA against stochastic simulations is the rate of immunity waning $\gamma$. In the limit $\gamma \to 0$, the SIRS model coincides with the SIR model, in the opposite limit $\gamma \to \infty$, with the SIS model, also known as the contact process \cite{anderson_may,ref_PA2,nikoloski}. At high $\gamma$ the process of loss of immunity by recovered individuals dominates over the infection that proceeds via $SI$ pairs. In this regime pairs are constantly changing type. Due to such randomization or homogenization of states  and to the randomness of the connections between the nodes in a RRG, the correlations are easily destroyed and an effective mixing of the population is achieved. Thus, it is expected that, in the limit $\gamma \gg 1$, the inclusion of short-range correlations in a deterministic model is sufficient for accurate description of the results of stochastic simulations.

In the limit $\gamma \ll 1$, the dynamics of the model is drastically different. In this regime the rate of immunity waning is smaller than the recovery rate $\delta$. As pointed out in the introduction, the corresponding region in the phase diagram is related in the epidemiological context with acute disease spread \cite{pre}. Recovery and  infection governed by $SI$ pairs dominate when $\gamma \ll 1$, and a RRG is crowded with recovered individuals while the overall number of infectives remains low. The density of $SI$ pairs through which the infection spreads is even lower. In this regime both stochastic effects due to finite size and any kind of "imperfections" in the structure of the RRG become important. Indeed, the standard PA is exact for tree-like structures where each node has exactly the same number of neighbors and there are no loops, the Bethe lattices. These infinite structures cannot be simulated on a computer. On the other hand, loops are inherent to all $d$-dimensional spatial structures used in computer simulations. Classic results of graph theory show that a particular realization of a RRG-$k$ will contain a large number of loops, of which the overwhelming majority are long (with respect to the average path length), so that locally the graph is essentially tree-like. One would expect then the qualitative prediction of the PA to perform well on RRGs, provided they are large enough. It is striking how the dynamics of the SIRS unfolding on RRGs appears to be so different from that predicted by the PA deterministic equations, revealing the subtle influence of long loops and stochastic effects whose role increases with decreasing $k$. 

We relate the observed suppression of oscillations also with the fact that in the PA model the oscillatory phase disappears for large $k$, where we could expect the PA to perform better. A similar observation of emergence and/or suppression of global oscillations in the qualitative and quantitative comparison of the results of Monte Carlo simulations on RRGs with the predictions of the  standard PA was reported in Refs. \cite{szabo1} and \cite{szabo2} for a spatial Rock-Scissors-Paper game, a system where three states cyclically dominate each other. In this model, for $k\geq 4$ the numerical solutions of the standard PA equations exhibit increasing amplitude oscillations of the state probabilities. Stochastic simulations of the model on RRGs show different qualitative behavior. For $k=3, 4$ the evolution tends towards a limit cycle and the growing spiral trajectories are observed on RRGs only for $k\geq6$. The six-node approximation predicts a limit cycle in good quantitative agreement with the simulations on a RRG-$3$ \cite{szabo1,szabo2}.

The standard PA considered in this paper can be applied straightforwardly for modeling the corresponding dynamics on networks with heterogeneous topologies. In this case, the parameter $k$ would stand for the first moment of the degree distribution. However, our results for random networks with homogeneous degree lead us to expect an even worse agreement with the numerical simulations on general heterogeneous networks. A number of recent studies have indicated that an improvement in the description of simple dynamics on more realistic network structures might be possible by introducing explicitly the degree dependence of the node and pair probabilities in the PA equations \cite{nikoloski,volz,pairhetero}.              

In conclusion, we have shown that two classes of models in population dynamics, an epidemic and a predator-prey model, with the coordination number $k$ as a parameter, exhibit similar phase diagrams in the PA. A distinctive feature of these diagrams is the presence of a small phase with stable oscillatory behavior in a weak driving regime of the active phase that is bounded by a curve associated with a supercritical Andronov-Hopf bifurcation. This oscillatory phase changes with $k$ and vanishes for large $k$ in both models.
The oscillatory phase has been associated with the possibility of unforced stable oscillations in predator-prey models and non-seasonal recurrent epidemics, however, even for RRGs, for which the PA is known to perform quite well, the results of stochastic simulations do not confirm the analytic prediction. In this regime the numerical results demonstrate that apart from finite size effects the macroscopic quasi-stationary state of the system is a nontrivial fixed point. 
The cyclic global behavior predicted by the PA is not observed in finite systems up to very large sizes. In order to capture the dynamics of these models in the weak driving
regime, an analytic description based on more elaborate approximation schemes than the PA must be considered, even when the underlying interaction network is a random graph. 
\begin{acknowledgement}
Financial support from the Foundation of the University of Lisbon 
and the Portuguese Foundation for Science and 
Technology (FCT) under contract POCTI/ISFL/2/618 is gratefully 
acknowledged. The first author (G.R.) was also supported by FCT under grant SFRH/BD/32164/2006 and by Calouste Gulbenkian Foundation under its Program "Stimulus for  Research". Additionally, one of the authors thanks Dr. I. Safonov for introducing her to new software tools. 
\end{acknowledgement}   



\begin{thebibliography}{}
%
%
\bibitem{murray}
J.~D.~Murray, \textit{Mathematical Biology I: An Introduction} (Springer-Verlag, New York, 2002).

\bibitem{anderson_may} 
R.~M.~Anderson and R.~M.~May, \textit{Infectious Diseases of Humans: Dynamics and Control} (Oxford University Press, Oxford, 1991).

\bibitem{contact_rev1}
\textit{The Geometry of Ecological Interactions: Simplifying Spatial Complexity}, edited by U.~Dieckmann, R.~Law, and J.~A.~J.~Metz (Cambridge University Press, Cambridge, 2000).

\bibitem{contact_rev2}
M.~J.~Keeling and K.~T.~D.~Eames, J. R. Soc. Interface \textbf{2}, 295 (2005).

\bibitem{ref_PA1}
H.~Matsuda, N.~Ogita, A.~Sasaki, and K.~Sato, Prog. Theor. Phys. \textbf{88}, 1035 (1992).

\bibitem{ref_PA2}
S.~Levin and R.~Durrett, Phil. Trans. R. Soc. Lond. B \textbf{351}, 1615 (1996). 

\bibitem{ref_PA3}
M.~J.~Keeling, D.~A.~Rand, and A.~J.~Morris, Proc. R. Soc. Lond. B \textbf{264}, 1149 (1997). 

\bibitem{ref_PA4}
M.~van~Baalen, in \textit{The Geometry of Ecological Interactions: Simplifying Spatial Complexity}, edited by U.~Dieckmann, R.~Law, and J.~A.~J.~Metz  (Cambridge University Press, Cambridge, 2000), p. 359. 

\bibitem{pre}
G.~Rozhnova and A.~Nunes, Phys. Rev. E \textbf{79}, 041922 (2009).

\bibitem{rand_morris_jerome1}
D.~A.~Rand, in \textit{Advanced Ecological Theory: Principles and Applications}, edited by J.~McGlade (Blackwell Science, Oxford, 1999), p. 100. 

\bibitem{rand_morris_jerome2}
A.~J.~Morris, PhD dissertation, University of Warwick, Coventry, UK, 1997. 

\bibitem{rand_morris_jerome3}
J.~Benoit, A.~Nunes, and M.~M.~Telo~da~Gama, Eur. Phys. J. B \textbf{50}, 177 (2006).

\bibitem{proc}
G.~Rozhnova and A.~Nunes, in \textit{Complex Sciences: Complex 2009, Part I, LNICST 4}, edited by J. Zhou (Springer Berlin Heidelberg, 2009), p. 792.

\bibitem{lebowitz}
J.~Joo and J.~L.~Lebowitz, Phys. Rev. E \textbf{70}, 036114 (2004).

\bibitem{tome1}
J.~E.~Satulovsky and T.~Tom{\'e}, Phys. Rev. E \textbf{49}, 5073 (1994).

\bibitem{tome2}
T.~Tom{\'e} and K.~C.~de~Carvalho, J. Phys. A: Math. Theor. \textbf{40}, 12901 (2007).

\bibitem{nikoloski}
Z.~Nikoloski, N.~Deo, and L.~Kucera, Complexus \textbf{3}, 169 (2006).

\bibitem{volz}
E.~Volz and L.~A.~Meyers, Proc. R. Soc. B \textbf{274}, 2925 (2007).

\bibitem{ourpre2}
G.~Rozhnova and A.~Nunes, Phys. Rev. E \textbf{80}, 051915 (2009).

\bibitem{bauchearn}
C.~T.~Bauch and D.~J.~D.~Earn, Proc. R. Soc. London, Ser. B {\bf 270}, 1573 (2003).

\bibitem{elton}
C.~S.~Elton, \textit{Voles, Mice and Lemmings: Problems in Population Dynamics} (Clarendon Press, Oxford, 1942).

\bibitem{exp_populations}
C.~B.~Huffaker, Hilgardia \textbf{27}, 343 (1958).

\bibitem{lynxhare}
J.~O.~Wolff, Ecological Monographs \textbf{50}, 111 (1980).

\bibitem{lotka}
A.~Lotka, J. Am. Chem. Soc. \textbf{42}, 1595 (1920).

\bibitem{volterra}
V.~Volterra, \textit{Le\c{c}ons sur la Th{\'e}orie Math{\'e}matique de la Lutte pour la Vie} (Gauthier-Villars, Paris, 1931).

\bibitem{lipowski}
A.~Lipowski, Phys. Rev. E \textbf{60}, 5179 (1999).

\bibitem{droz}
T.~Antal and M.~Droz, Phys. Rev. E \textbf{63}, 056119 (2001).

\bibitem{mobilia}
M.~Mobilia, I.~T.~Georgiev, and U.~C.~T{\"a}uber, J. Stat. Phys. \textbf{128}, 447 (2007).

\bibitem{alava}
M.~Peltom{\"a}ki, M.~Rost, and M.~Alava, Phys. Rev. E \textbf{78}, 050903(R) (2008).

\bibitem{predpreyparameters}
S.~Boutin, Wildl. Res. \textbf{22}, 89 (1995).

\bibitem{gillespie1}
A.~B.~Bortz, M.~H.~Kalos, and J.~L.~Lebowitz, J. Comput. Phys. \textbf{17}, 10 (1975).

\bibitem{gillespie2}
D.~T.~Gillespie, J. Comput. Phys. \textbf{22}, 403 (1976).

\bibitem{szabo1}
G.~Szab\'o, A.~Szolnoki, and R.~Izs\'ak, J. Phys. A \textbf{37}, 2599 (2004).

\bibitem{szabo2}
A.~Szolnoki and G.~Szab\'o, Phys. Rev. E \textbf{70}, 037102 (2004).

\bibitem{pairhetero}
E.~Pugliese and C.~Castellano, e-print arXiv:0903.5489.

\end{thebibliography}
\end{document}